\documentclass[11pt]{article}
\usepackage{amsmath,epsfig,graphicx,amssymb}
\usepackage[margin=0.7in]{geometry}
\newcommand{\beq}{\begin{equation}}
\newcommand{\eeq}{\end{equation}}
\newcommand{\ben}{\begin{eqnarray}}
\newcommand{\een}{\end{eqnarray}}
\newcommand{\bes}{\begin{subequations}}
\newcommand{\ees}{\end{subequations}}
\newcommand{\bFig}{\begin{figure}}
\newcommand{\eFig}{\end{figure}}

\date{}
\begin{document}

\title{Hilbert Space Theory of Classical Electrodynamics}
\author{A. K. Rajagopal\footnote{attipat.rajagopal@gmail.com}\\
(a) Inspire Institute Inc., Alexandria, Virginia,22303, USA\\
(b) Harish-Chandra Research Institute, Chhatnag Road,Jhunsi, Allahabad, 211 019, India\\
(c) Institute of Mathematical Sciences, C. I.T. Campus, Tharamani, Chennai, 600 113, India\\
and\\
Partha Ghose\footnote{partha.ghose@gmail.com} \\
Centre for Astroparticle Physics and Space Science (CAPSS),\\Bose Institute, \\ Block EN, Sector V, Salt Lake, Kolkata 700 091, India.}
\maketitle

\begin{abstract}
Classical electrodynamics is reformulated in terms of wave functions in the classical phase space of electrodynamics, following the Koopman-von Neumann-Sudarshan prescription for classical mechanics on Hilbert spaces {\em sans} the superselection rule which prohibits interference effects in classical mechanics. This is accomplished by transforming from a set of commuting observables in one Hilbert space to another set of commuting observables in a larger Hilbert space.  This is necessary to clarify the theoretical basis of much recent work on quantum-like features exhibited by classical optics. Furthermore, following Bondar et al ({\em Phys.Rev. A} {\bf 88}, 052108, (2013)), it is pointed out that quantum processes that preserve the positivity or nonpositivity of the Wigner function can be implemented by classical optics. This may be useful in interpreting quantum information processing in terms of classical optics.
\end{abstract}
\section{Introduction}
Much recent work have unexpectedly revealed that classical optics displays some quantum-like behaviour like entanglement, as originally predicted by Spreeuw \cite{sp} and independently by Ghose and Samal \cite{g}. This emerging field has been reviewed by Ghose and Mukherjee \cite{ghose} with emphasis on the Hilbert space structure of classical polarization optics. Simon et al \cite{simon} have based their analysis of the classical optics framework in the mathematical structure of Mueller matrices that are employed therein. Their analysis suggests that it is the kinematic structure of classical optics, namely its Hilbert space structure, that gives rise to the surprisingly quantum-like features in classical optics. The words ``entanglement'' and ``Bell violation'' are correlation features of quantum mechanics. Qian and Eberly \cite{q1,q2} have shown that, without contradicting quantum mechanics,  classical non-deterministic fields also provide a natural basis for entanglement and Bell analyses. Surprisingly, these fields are not eliminated by the Clauser-Horne-Shimony-Holt Bell violation tests as viable alternatives to quantum theory. They proposed an experimental setup for its verification. More recently, Aiello et al \cite{aiello} have attempted to develop a unified theory of different kinds of light beams exhibiting `classical entanglement' and indicated several possible extensions of the idea, and Pereira et al \cite{pereira} have investigated the difference between classical and quantum inseparability in the case of spin-orbit coupling through violations of an intensity based CHSH inequality. There is an urgent need therefore to understand more comprehensively the origin of such quantumness in patently classical phenomena. It should be pointed out that exclusive quantum effects, such as the Aharonov-Bohm effect, are not obtainable in these theories as they pertain to non-commuting operators in Hilbert space whereas classical effects are associated with commuting operators. It is therefore important to find the difference between the KvNS Hilbert space and the QM Hilbert space, which is addressed here.

We will show in this paper that the answer lies in the classic works of Koopman \cite{K}, von Neumann \cite{vN} and later others \cite{kvn1,kvn2,kvn3,kvn4,kvn5,mauro,kvn6,kvn7,kvn8,kvn9,kvn10,kvn11,kvn12,kvn13, kvn14}. Sudarshan \cite{sud} developed a complete theory of Classical Mechanics (CM) based on Hilbert spaces associated with commuting hermitian operators as observables (hereinafter referred to as the KvNS theory). There are some recent papers that extend the KvNS formalism \cite{mauro,bond,bond2}, but they are not our concern here. The operators that correspond to position and momentum in the KvNS formalism are taken to commute with each other. The formalism introduces additional canonically conjugate operators $\hat{\lambda}_q,\hat{\lambda}_p$ which are deemed `unobservable'. They have the properties $[\hat{q}, \hat{\lambda}_q]=i=[\hat{p}, \hat{\lambda}_p]$ and all others commuting among themselves. Also, the superposition of states occurs in ensembles of classical states which do not correspond to macroscopic Schr\"{o}dinger-cat like states. Even though the KvNS theory is a reformulation of Classical Mechanics (CM) on a Hilbert space, it is not known as yet to provide insight into how CM can emerge from QM, or vice versa. 

Our focus in this paper is on classical electrodynamics because it is in this area of research that the quantum-like signatures mentioned above have so far been displayed in the literature. The paper is divided into four sections. Section 2 describes the KvNS theory with two subsections for making the paper self-contained. The first one briefly reviews the salient features of the original KvNS theory of classical mechanics, and the second subsection develops the full KvNS theory of classical electrodynamics. This is followed by Section 3 in which the original commuting set of operators is transformed to another commuting set which enabled Bondar et al \cite{bond} to deduce that the KvNS wave function is a probability amplitude for a quantum particle at a certain point of the classical phase space, implying that certain quantum processes are implementable by classical means. The final section 4 summarizes the work, thus rounding off this interesting and important observation that the Hilbert Space framework in some general sense covers both quantum and classical features with different implications. This does not in any way mean that QM and CM are equivalent, but explains why one should expect to witness QM-like features in the classical framework when viewed in a generalized Hilbert Space theory such as KvNS.

\section{KvNS Theory}
\subsection{Classical Mechanics}
The basic idea of the Koopman theory \cite{K} lies in the classical phase space description of statistical mechanics.  For simplicity of presentation here, we consider a single-particle system described by a Hamiltonian $H(p,q)$ in 2-dimensional classical phase space consisting of commuting position and momentum variables, and the distribution function $f(q,p;t)$. The classical Liouville equation obeyed by this distribution is given by
\beq
\frac{\partial f(q,p;t)}{\partial t} = \left(\frac{\partial H}{\partial q}\frac{\partial}{\partial p} - \frac{\partial H}{\partial p}\frac{\partial}{\partial q}\right)f(q,p;t) \equiv -i\hat{L}f(q,p;t)
\eeq 				
This was immediately followed up by von Neumann \cite{vN}, and the two authors postulated that this can be looked upon as arising from classical square integrable wave functions $\psi(q,p;t)$ in the Hilbert space of classical phase space variables $\psi(q,p;t)$ obeying the time-development equation										
\beq
\frac{\partial \psi(q,p;t)}{\partial t} = \left(\frac{\partial H}{\partial q}\frac{\partial}{\partial p} - \frac{\partial H}{\partial p}\frac{\partial}{\partial q}    \right)\psi(q,p;t)= -i\hat{L}\psi(q,p;t).\label{2}
\eeq 					
Its complex conjugate is given by
\beq
\frac{\partial \psi^*(q,p;t)}{\partial t} = \left(\frac{\partial H}{\partial q}\frac{\partial}{\partial p} - \frac{\partial H}{\partial p}\frac{\partial}{\partial q}    \right)\psi^*(q,p;t)=-i\hat{L}\psi^*(q,p;t).\label{3}
\eeq 
The Hilbert space is equipped with the scalar product
\beq
\langle \phi|\psi\rangle = \int dp\, dq\, \phi^*(q,p)\psi(q,p).
\eeq				
The classical coordinates and momenta are here represented by commuting operators. This Hilbert space is twice the size of the usual case and the integrability condition is thus given by
\beq
||\psi||^2=\int\int dq\, dp\, \psi^*(q,p;t)\psi(q,p;t) = N.
\eeq
The integration goes over all $q,p$ from $-\infty$ to $+\infty$ . 
We will use dimensionless variables in this development and identify the density in phase space by
\beq
\psi^*(q,p;t)\psi(q,p;t)=|\psi(q,p;t)|^2 \equiv \rho(q,p;t)\label{rho}
\eeq
From eqs.(\ref{2},\ref{3}), we deduce the equation obeyed by the density defined by eq.(\ref{rho}),
\beq
\frac{\partial \rho(q,p;t)}{\partial t} = \left(\frac{\partial H}{\partial q}\frac{\partial}{\partial p} - \frac{\partial H}{\partial p}\frac{\partial}{\partial q}    \right)\rho(q,p;t) \Rightarrow i\frac{\partial \rho}{\partial t} = \hat{L}\rho.\label{5}
\eeq
This is a derivation of the standard Liouville equation, starting from the postulated eqn. (\ref{2}) for the classical wave function $\psi(q,p;t)$. It shows that the dynamics of $\rho$, the probability density, can be recovered from the dynamics of the underlying wave function $\psi$ and $\psi^*$. 

The doubling of the variables in the Liouvillean theory is natural and has been used elsewhere. It is also used within quantum field theory as well as in the density matrix theory of quantum mechanics. A classical mechanics was later developed by Sudarshan \cite{sud} as a `hidden variables' quantum theory with only the absolute values of the $\psi$s as relevant, i.e. $\psi = \sqrt{\rho}$. In other words, in Sudarshan's theory a `superselection rule' operates that makes all transitions between quantum wave functions with different phases unobservable, resulting in classical mechanics as a `hidden variable' quantum theory. This construction of Sudarshan is to be contrasted with that of Koopman and von Neumann.  

To elucidate how the KvN formalism also triggers a `superselection rule' that decouples the phase and the amplitude of the KvNS wave function, consider the simple case of a free particle Hamiltonian $H = p^2/2m$. Then $q$ is a cyclic coordinate and $p$ is a constant of motion and $q =(p/m)t$ + constant. However, Mauro \cite{mauro} shows that a suitable representation of the phase space reveals the phase feature for free particles as well, which we describe now. The Liouville equation is
\beq
\frac{\partial \rho(\varphi)}{\partial t} = -\frac{\partial}{\partial q}\left(\frac{p}{m}\rho(\varphi)\right)\label{L1}
\eeq
Thus, although the Liouville equation $d\rho/dt=0$ holds, this continuity equation features only the square of the amplitude $\sqrt{\rho}$ of the wave function but not its phase $S$ which can be defined by writing $\psi=\sqrt{\rho}\, {\rm exp}(iS)$. This can also be seen by insering this expression for $\psi$ in equations (\ref{2}) and (\ref{3}) and separating the real and imaginary parts to get
\beq
i\frac{\partial \sqrt{\rho}}{\partial t} = \hat{L}\sqrt{\rho},\,\,\,\,i\frac{\partial S}{\partial t} = \hat{L}S \label{L2}.
\eeq
This shows that the amplitude and phase evolve independently. This is a result of a `superselection rule', and it is possible to work only with $\sqrt{\rho}$.

The commuting $\hat{q},\hat{p}$ operators have continuous spectra from $-\infty$ to $+\infty$, 
\beq
\hat{q}|q,p\rangle = q|q,p\rangle,\,\,\,\,\hat{p}|q,p\rangle = p|q,p\rangle,\label{a1}
\eeq
with $\{|q,p\rangle\}$ spanning the KvN Hilbert space, being orthonormal and a complete set:
\ben
\langle q^{'},p^{'}|q^{''},p^{''}\rangle &=& \delta(q^{'} - q^{''})\delta(p^{'} - p^{''}),\\
\int \int dq\,dp\,|q,p\rangle\langle q,p|&=&1.
\een
The operators $-i\partial_q$ and $-i\partial_p$ are hermitian in this representation:
\ben
\langle q^{'}, p^{'}|-i\partial_q|\psi\rangle = -i\partial_{q^{'}}\langle q^{'}, p^{'}|\psi\rangle,\label{b1}\\
\langle q^{'}, p^{'}|-i\partial_p|\psi\rangle = -i\partial_{p^{'}}\langle q^{'}, p^{'}|\psi\rangle. \label{b2}
\een
Hence it follows from (\ref{a1}), (\ref{b1}) and (\ref{b2}) that
\ben
\langle q^{'}, p^{'}|[\hat{q},-i\partial_q]|\psi\rangle = \langle q^{'}, p^{'}|i|\psi\rangle,\\
\langle q^{'}, p^{'}|[\hat{p},-i\partial_p|\psi\rangle = \langle q^{'}, p^{'}|i|\psi\rangle, 
\een
and all other commutators vanish. Thus, there are two more commuting operators in the KvN Hilbert space, $\hat{\lambda}_q = -i\partial_q$ and $\hat{\lambda}_p = -i\partial_p$, with the properties $[p,\hat{\lambda}_p] =i=[q,\hat{\lambda}_q ]$. The Hilbert space can be described by any pair of commuting operators, and there are 4 choices: $(q,p), (q,\lambda_p), (\lambda_q,p)$ and $(\lambda_q,\lambda_p)$. 
The classical Liouville equation for the classical state is written as
\beq
\frac{\partial}{\partial t}|\psi(t)\rangle = \left(\partial_q H\hat{\lambda}_p - \partial_p H\hat{\lambda}_q \right)|\psi(t)\rangle.
\eeq

Let us now make a unitary transformation from the commuting variables $(q,p)$ to the commuting variables $(q,\lambda_p)$ \cite{mauro} such that
\beq
\hat{q}|q,\lambda_p\rangle = q|q,\lambda_p\rangle,\,\, \hat{\lambda}_p|q,\lambda_p\rangle = \lambda_p|q,\lambda_p\rangle.
\eeq 
The transformation equations are therefore
\beq
-i\frac{\partial}{\partial p^{'}} \langle q^{'},p^{'}|q,\lambda_p\rangle = \lambda_p \langle q^{'},p^{'}|q,\lambda_p\rangle
\eeq
so that
\beq
\langle q^{'},p^{'}|q,\lambda_p\rangle = \frac{1}{\sqrt{2\pi}}\delta (q -q^{'})\,{\rm exp}\left(ip^{'}\lambda_p\right).\label{tr}
\eeq
Thus, $\{|q,\lambda_p\rangle\}$ form a complete orthonormal set related to the $\{|q,p\rangle \}$ complete set by the transformation function (\ref{tr}). Hence,
\beq
\langle q, \lambda_p|\psi\rangle = \int\int dq^{'} dp \langle q, \lambda_p|q^{'},p\rangle\langle q^{'},p|\psi\rangle =  \frac{1}{\sqrt{2\pi}} \int dp\, {\rm exp}\left(ip \lambda_p\right)\langle q,p|\psi\rangle,
\eeq
or equivalently,
\ben
\psi(q,\lambda_p,t) &=& \frac{1}{\sqrt{2\pi}} \int dp\, {\rm exp}\left(ip \lambda_p\right)\psi(q,p,t),\label{psi}.
\een

Let us consider a particle in an external potential $V(q)$. The Hamiltonian is $H = p^2/2 + V(q)$, and we obtain the Liouville equation for the classical wavefunction
\beq
i\frac{\partial}{\partial t}\psi(q,\lambda_p,t)=\left(\frac{\partial}{\partial q}\frac{\partial}{\partial \lambda_p} -V^{'}(q)\lambda_p \right)\psi(q,\lambda_p,t)\label{C1}
\eeq
and its complex conjugate
\beq
i\frac{\partial}{\partial t}\psi^*(q,\lambda_p,t)=-\left(\frac{\partial}{\partial q}\frac{\partial}{\partial \lambda_p} -V^{'}(q)\lambda_p \right)\psi^*(q,\lambda_p,t),\label{C2}
\eeq
from which we deduce the density-current relationship
\ben
\frac{\partial}{\partial t}\rho(q,\lambda_p,t) &=& -J(q,\lambda_p,t),\nonumber\\
J(q,\lambda_p,t) &=& i\left(\psi^*\frac{\partial}{\partial q}\frac{\partial}{\partial \lambda_p}\psi -\frac{\partial}{\partial q}\frac{\partial}{\partial \lambda_p}\psi^* \psi\right) \label{L3}
\een
If we now write
\beq
\psi(q,\lambda_p,t)= \sqrt{\rho(q,\lambda_p,t)}\,{\rm exp}\,iS(q,\lambda_p,t),\label{L4}
\eeq
we see that the time derivatives of density and amplitude get connected unlike in the (q,p) representation, but within the classical Hilbert space theory. This shows that in the new alternate representation of the classical KvN Hilbert space, phase features are present. For a free particle considered in equations (\ref{L1}, \ref{L2}), $V(q)= 0$, and the $(q,\lambda_p)$ representation now exhibits a phase in its KvN wave functions (eqns.(\ref{L3}, \ref{L4}).

If one wishes this phase of the wave function to be unobservable, like in the $(q,p)$ representation, one has to invoke a superselection rule, prohibiting superpositions of states to be written. This can be done by requiring that only the positive square root $\sqrt{\rho(\varphi)}$ of the probability density is physically relevant \cite{sud}. This can be understood in terms of the structure of Hilbert spaces in the following way. The absolute phase of a state is not measurable either in quantum mechanics or in classical optics. This is why all states (vectors in Hilbert space) which differ only by a phase are identified to get `rays' or `projective rays'. The space of rays is therefore $CP = H/U(1)$. Examples are the Bloch sphere in quantum mechanics and the Poincare sphere in classical polarization optics. Relative phases, however, remain observable in both quantum mechanics and classical optics, but not in classical mechanics. To avoid relative phases in classical mechanics, a further projection or identification is required, namely, all states that differ by relative phases must also be identified. Hence the space in classical mechanics must be the quotient space $CP^* = CP/U(1)$. This results in the `superselection rule' operating on CP*, preventing superposition of states and interference. An example would be the identification/projection of all circles on a sphere $S^2$ to a single circle $S^1$.

A few remarks are in order here. First, the $(q,p)$ and $(q, \lambda_p)$ representations are made physically inequivalent by changing the observables from functions of the commuting set $(q,p)$ to functions of the commuting set $(q, \lambda_p)$ \cite{mauro}. This corresponds to an enlargement of the space $CP^*$ to $CP$. The $CP$ space allows coherent superpositions of states and renders their relative phases observable. Second, if one imposes the superselection condition on the $(q, \lambda_p)$ representation, it becomes equivalent to the $(q,p)$ representation. This is reflected in the reduction of the $CP$ space to $CP^*$. Third, the crucial differences between a classical Hilbert space theory with phases and a quantum theory lie in the commutativity of all observables in the classical theory as well as in the different dynamical evolution equations in the two cases. Finally, there are other possible representations as indicated above, and they could presumably be of some importance in the grey region at the interface of classical and quantum mechanics. 

Since a classical electromagnetic field is formally an infinite collection of classical harmonic oscillators and shows interference effects, it should be possible to develop a KvN type theory of such fields by using a projective Hilbert space $CP$ so that no superselection rule operates. Such a field will have coherence properties like those of quantum mechanics except those specifically associated with quantization. This is what we do in the next section. 

\subsection{Classical Electrodynamics}
The classical electromagnetic field is specified by the complex valued electric and magnetic fields $\vec{E}(\vec{x},t)$ and $\vec{B}(\vec{x},t)$ at every space-time point. Hence, let the classical `wave function' of the electromagnetic field be a six-component column vector
\beq
\psi(E_i,B_i)= \left(\begin{array}{c}
E_x \\E_y\\E_z\\-B_x\\-B_y\\-B_z
\end{array} \right)  
\eeq
and its dual $\psi^\dagger (E_i,B_i) = \left(E^*_x,E^*_y, E ^*_z, -B^*_x, -B^*_y, -B^*_z\right)$. These wave functions are functions of all the phase space variables, which are all the components of the fields. For free fields there is a constraint due to gauge invariance, namely $\vec{E}.\vec{B}=0$ which can be taken care by choosing the unit polarization vectors of the two fields $\vec{e}_E$ and $\vec{e}_B$ to be orthogonal, $\vec{e}_E.\vec{e}_B=0$.
There are two more kinematic constraints, namely ${\rm div} \vec{E}=0$ and ${\rm div} \vec{B}=0$. 
Denoting the set of complex dynamical variables $\{E_i, B_i\}$ by $\varphi$, one can write superpositions $\alpha \psi(\varphi) + \beta \phi(\varphi)$ with arbitrary complex coefficients $\alpha$ and $\beta$, and define inner products
\beq
\langle \phi|\psi\rangle = \int \Pi_i dE_i\Pi_j dB_j \delta(\vec{E}.\vec{B})\langle \phi|E_i,B_i\rangle\langle E_i,B_i|\psi\rangle =  \int d\varphi\, \phi^\dagger(\varphi)\psi(\varphi)
\eeq 
where the integration measure is the phase space volume $d\varphi=\Pi_i dE_i \Pi_j dB_j\delta(\vec{E}.\vec{B})$. 
The scalar product is given by 
\beq
||\psi||^2=\langle \psi|\psi\rangle = \int d\varphi\, \psi^\dagger (\varphi) \psi(\varphi) = \int d\varphi\,\rho(\varphi)
\eeq
with $\rho(\varphi) =\psi^\dagger (\varphi) \psi(\varphi)= \sum_i(E_i^*E_i + B_i^*B_i)$ ($i=x,y,z$), the energy density of the field which is a collection of three independent harmonic oscillators with the constraint $\vec{E}.\vec{B}=0$ which reduces the effective number to two. Normalized by the total energy, this gives the probability density in phase space if one treats the magnetic fields $\{B_i\}$ as the cordinates and the electric fields $\{E_i\}$ as the conjugate momenta such that $\vec{E}.\vec{B}=0$, leading to two orthogonal sets of fields $(E_x,0,0,0,B_y,0)$ and $(0,E_y,0,B_x,0,0)$, and the third one orthogonal to these two does not exist in the free field case as there are no charges and currents. These constitute the two oscillators associated with free EM fields. The $\psi(\varphi)$ are square integrable functions and span a complex Hilbert space, and $\psi^\dagger(\varphi)$ are their duals. In classical theory $\vec{E}$ and $\vec{B}$ commute. Using units in which $\epsilon_0 = \mu_0 = c =1$, let the equations of motion for the wave functions be
\ben
\frac{\partial}{\partial t}\psi(\varphi,t) &=& -i\hat{L}\psi(\varphi,t) = -\sum_i\beta_i \partial_i\psi(\varphi,t),\label{eq1}\\
\frac{\partial}{\partial t}\psi(\varphi,t)^\dagger &=& -i\psi(\varphi,t)^\dagger\hat{L} =-\sum_i\partial_i\psi(\varphi,t)^\dagger\beta_i,\label{eq2} 
\een
where $\partial_i \equiv \partial/\partial {x}_i$ and $\hat{L}=-i\sum_i\beta_i \partial_i$ is the Liouvillian operator and $\beta_i^\dagger = \beta_i$ are hermitian operators with the following matrix representations:
\vskip 0.1in
$\beta_x =\left(\begin{matrix}
0&0&0&0&0&0\\0&0&0&0&0&-1\\0&0&0&0&1&0\\
0&0&0&0&0&0\\0&0&1&0&0&0\\0&-1&0&0&0&0
\end{matrix}\right)$,\,\,\,$\beta_y =\left(\begin{matrix}
0&0&0&0&0&1\\0&0&0&0&0&0\\0&0&0&-1&0&0\\
0&0&-1&0&0&0\\0&0&0&0&0&0\\1&0&0&0&0&0
\end{matrix}\right)$,\,\,\,$\beta_z =\left(\begin{matrix}
0&0&0&0&-1&0\\0&0&0&1&0&0\\0&0&0&0&0&0\\
0&1&0&0&0&0\\-1&0&0&0&0&0\\0&0&0&0&0&0
\end{matrix}\right)$
\vskip 0.1in
Hence
\vskip 0.1in
-$\sum_i\beta_i\partial_i =\left(\begin{matrix}
0&0&0&0&\partial_z&-\partial_y\\0&0&0&-\partial_z&0&\partial_x\\0&0&0&\partial_y&-\partial_x&0\\
0&-\partial_z&\partial_y&0&0&0\\\partial_z&0&-\partial_x&0&0&0\\-\partial_y&\partial_x&0&0&0&0
\end{matrix}\right)$
\vskip 0.1in
These hermitian matrices satisfy the algebra
\beq
[\beta_i,\beta_j] = \epsilon_{ijk}\beta_k.
\eeq
It is straightforward to verify that the equations (\ref{eq1}) and (\ref{eq2}) encode the Maxwell equations
\beq
\dot{\vec{E}} = {\rm curl}\vec{B},\,\,\,\,\dot{\vec{B}} = -{\rm curl}\vec{E}.
\eeq
Multiplying the first equation by $\psi(\varphi,t)^\dagger$ from the left and the second equation by $\psi(\varphi,t)$ from the right and adding them, one gets the continuity equation
\ben
\frac{\partial \rho(\varphi,t)}{\partial t} &=& - 2i\psi(\varphi,t)^\dagger\hat{L}\psi(\varphi,t) = - \sum_i\partial_i S_i(\varphi,t),\nonumber\\
S_i(\varphi,t) &=& \psi^\dagger(\varphi,t) \beta_i \psi(\varphi,t).\label{cont}
\een
It is easily verified that $S_i=(\vec{E}\times \vec{B})_i$ are components of the Poynting vector. This is the Liouville equation for the phase density.

In order to go over to a KvN type Hilbert space theory in $CP$ space, 
let us now make a unitary transformation from the commuting variables $\varphi=(B,E)$ to the commuting variables $\chi= (B,\lambda_E)$ in the KvN Hilbert space such that
\beq
\hat{B_i}|B,\lambda_E\rangle = B_i|B,\lambda_E\rangle,\,\, \hat{\lambda}_{E_i}|B,\lambda_E\rangle = \lambda_{E_i}|B,\lambda_E\rangle
\eeq
with the properties $\hat{\lambda}_{E_i} = -i\partial_{E_i}$, $[E_i,\hat{\lambda}_{E_i}] =i$,  and with the transformation
\beq
\langle B^{'},E^{'}|B,\lambda_E\rangle = \frac{1}{(2\pi)^{3/2}}\Pi_i\delta (B_i -B_i^{'})\,{\rm exp}\left(i\sum_iE_i^{'}\lambda_{E_i}\right).\label{tr2}
\eeq
Following essentially the same procedure as defined in the previous section, one obtains
\ben
\psi(\chi,t) &=& \frac{1}{(2\pi)^{3/2}} \int d^3 E\, {\rm exp}\left(i\sum_iE_i \lambda_{E_i}\right)\psi(\varphi,t).
\een
Let us introduce the corresponding hermitian operator $\hat{\lambda}_{B_i} = -i\partial_{B_i}$ with the properties $\hat{\lambda}_{B_i}|B,\lambda_E\rangle = \lambda_{B_i}|B,\lambda_E\rangle$, $[B_i,\hat{\lambda}_{B_i}] =i$.
The Liouville equations for the classical wave function and its adjoint (sum on repeated indices assumed) in the new variables are 
\ben
i\frac{\partial }{\partial t}\psi(\chi,t) &=& \left(\frac{\partial}{\partial B_i^*}\frac{\partial}{\partial \lambda_{E_i}} - B_i^*\lambda_{E_i} \right)\psi(\chi,t),\label{CEM1}\\
i\frac{\partial }{\partial t}\psi^*(\chi,t) &=& -\left(\frac{\partial}{\partial B_i^*}\frac{\partial}{\partial \lambda_{E_i}} - B_i^*\lambda_{E_i} \right)\psi^*(\chi,t).\label{CEM2}
\een
It follows from these equations that
\ben
\frac{\partial }{\partial t}\rho(\chi,t) &=& -J(\chi,t),\\
J(\chi,t) &=& i\left(\psi^*\frac{\partial}{\partial B_i^*}\frac{\partial}{\partial \lambda_{E_i}}\psi -\frac{\partial}{\partial B_i^*}\frac{\partial}{\partial 
\lambda_{E_i}}\psi^* \psi\right).
\een
Writing $\psi(\chi,t)$ in the form
\beq
\psi(\chi,t)= \sqrt{\rho(\chi,t)}\,{\rm exp}\,iS(B_i^*,\lambda_{E_i},t)
\eeq
makes it clear that the time derivatives of the density and amplitude get connected within the classical Hilbert space theory of the electromagnetic field. This shows that phase features are preserved in the $(B_i, \lambda_{E_i})$ representation of the classical KvN Hilbert space of the electromagnetic field. This completes the construction of a KvNS theory of the classical electromagnetic field in Hilbert space. A Hilbert space structure is sufficient to permit entanglement (in the sense of non-separability) by virtue of the Schmidt theorem \cite{schmidt} which predates quantum mechanics, and hence Bell violations \cite{q1, q2}.
 
In this context the work of Bondar et al \cite{bond}, which we will summarize in the next seaction, is of particular interest. It shows that the KvNS wave function is the probability amplitude for a quantum particle at a certain point of the classical phase space. It clarifies the meaning of a Hilbert space in relation to classical features of the quantum structure. This requires a change of representation from $\{q,p\}$ to $\{q,\lambda_p\}$. 

\section{The Significance of the Wigner Function}
Recently Bondar et al \cite{bond} have reformulated the KvN theory following a Wigner function-like formulation. The significance of the Wigner function in quantum phase space has so far been that it mimics the classical distribution function but with characteristic quantum features imbedded in it. Specifically, this phase space distribution is not positive everywhere, the quantumness being conventionally attributed to its negative features. What Bondar and his coworkers have shown is that the Wigner function is a wave function rather than a distribution function, and hence is `a probability amplitude for the quantum particle to be at a certain point of the classical phase space'. It reduces to the KvNS wave function rather than to a classical distribution function in the classical limit. Since wave functions can be negative, the `mystery' of the negativity of the Wigner function disappears. The essential quantumness of a process lies not in the negativity of the Wigner function but in the distinctiveness of quantum processes to make transitions from positivity to nonpositivity and vice versa compared to negativity and positivity preserving processes only in classical physics.

We give now a summary of this formulation for completeness.  
The point of departure from KvN and \cite{mauro} in this formulation is to define new momentum and coordinate operators obeying a quantum-like commutation relation
\beq
[\hat{q}_Q, \hat{p}_Q] =i\hbar \kappa,\,\,0\leq\kappa\leq 1 \label{a}
\eeq
These operators are defined in terms of the classical operators
\beq
\hat{q}_Q = \hat{q} - \frac{1}{2}\hbar \kappa \hat{\lambda}_p, \,\,\hat{p}_Q = \hat{p} + \frac{1}{2}\hbar \kappa \hat{\lambda}_q,\label{b}
\eeq
The extra information beyond the KvN Hilbert space structure is contained in the operators (\ref{a}, \ref{b}) which obey the Ehrenfest theorems, from which a unique Hamiltonian-like operator $\hat{H}_{QC}$ can be defined such that
\ben
i\hbar \frac{d}{dt} |\Psi_\kappa (t)\rangle &=& \hat{H}_{QC}|\Psi_\kappa (t)\rangle,\label{c}\\
\hat{H}_{QC} &=& \hbar \hat{p}\hat{\lambda}_q + \frac{1}{\kappa}\left(U(\hat{q} - \frac{1}{2}\hbar \kappa \hat{\lambda}_p) - U(\hat{q} + \frac{1}{2}\hbar \kappa \hat{\lambda}_p)\right).
\een
The states $\{|\Psi_\kappa (t)\rangle\}$ are a complete orthonormal set of solutions of eq.(\ref{c}).
The parameter $\kappa$ denotes the quantum structure if it is equal to 1 and classical nature if it is set equal to 0. In fact, in the classical limit $\hat{H}_{QC}$ goes to $\hbar \hat{L}= \hbar \left(\hat{p}\hat{\lambda}_q - U^{'}(q)\hat{\lambda}_p \right)$. The important point that emerges from this formulation is that the unified wave function $|\Psi_\kappa (t)\rangle$ defined by eq.(\ref{c}) in the qp-representation is proportional to the Wigner function $W(q,p)$,
\ben
\langle qp|\Psi_\kappa (t)\rangle &=& \sqrt{2\pi\hbar\kappa}W(q,p),\nonumber\\
W(q,p) &=& \int \frac{dy}{2\pi\hbar\kappa}\rho_\kappa \left(q - \frac{y}{2}, q + \frac{y}{2}\right){\rm exp}\left(\frac{ipy}{\hbar\kappa}\right)\label{d}
\een
Here $\rho_\kappa \left(q - \frac{y}{2}, q + \frac{y}{2}\right)$ is the density matrix and is proportional to $\langle q\lambda_p|\Psi_\kappa\rangle$ and the transformation from the $q\lambda_p$ to the $qp$ representation gives the expression in eq.(\ref{d}). The normalization of the states $\langle\Psi_\kappa|\Psi_\kappa\rangle =1$ implies that the density matrix corresponds to a pure state $\hat{\rho}_\kappa^2 = \hat{\rho}_\kappa$. Thus the Wigner function of a pure state maps the quantum wave function into a corresponding KvN classical wave function rather than to a classical phase space distribution, as was originally thought by KvN! Thus, what has been discovered by this method of Bondar et al is the important fact that the KvNS wave function is the probability amplitude for a quantum particle at a certain point of the classical phase space, which implies that certain positivity- and nonpositivity-preserving quantum processes are implementable by classical means. This explains why classical optical methods may simulate some quantum gates \cite{sp,ghose}.

\section{Concluding Remarks}
The basic motivation of this paper has been to clarify the theoretical basis of quantum-like features such as `entanglement' and `Bell violation' exhibited by classical optics. This has been done by formulating classical electrodynamics as a complete phase space theory based on Hilbert spaces. The guiding principles were laid down long ago by Koopman, von Neumann and Sudarshan who developed a complete theory of classical mechanics based on the phase space description of statistical mechanics. The wave functions in KvNS theory are square integrable functions which span a Hilbert space, just as quantum mechanical wave functions do, with the difference that a superselection rule has to be invoked in the case of classical mechanics to prevent coherent superpositions. No such rule need be invoked in the case of classical electrodynamics. This demystifies the occurrence of interference, `entanglement' and `Bell violation' that have already been observed in classical optics. Viewed in this way, the main difference between classical  and quantum theories lies in the absence of non-commuting operators in the former. In other words, it is not only the projective Hilbert space structure of quantum mechanics but also its non-commutative operator structure that distinguishes it from classical optics. 

The recent work of Bondar and coworkers is also briefly sketched in order to bring out the connection of these KvNS wave functions and Wigner functions, showing that the latter reduce to the former in the classical limit and not to classical distribution functions in phase space, as was originally believed. This is significant in showing that the Wigner function is a probability amplitude rather than a probability distribution for a quantum particle at a certain point of the classical phase space, and amplitudes can be negative as well as positive. This has the important implication that certain quantum processes, namely those which preserve the positivity or nonpositivity of the Wigner function, are implementable by classical means. This may be of practical importance for understanding quantum information processing, based on Hibert space theory of classical optics. 
\section{Acknowledgement}
PG thanks the National Academy of Sciences, India for the grant of a Senior Scientist Platinum Jubilee Fellowship that enabled this work to be undertaken.


\begin{thebibliography}{0}
\bibitem{sp}
R. J. C. Spreeuw, {\em Found. of Phys.} {\bf 28} 361 (1998); {\em Phys. Rev. A} {\bf 63}, 062302 (2001).
\bibitem{g}
P. Ghose \& M. K. Samal, arXiv:quant-ph/0111119v1 22Nov 2001.
\bibitem{ghose}
Partha Ghose and  A. Mukherjee, {\em Rev. in Theoret. Sc.} {\bf 2}, 1-14 (2014); arXiv: 1308.6154.
\bibitem{simon}
B. N. Simon et al, {\em Phys. Rev. Letts.} {\bf 104}, 023901 (2010). 
\bibitem{q1}
X. F. Qian and J. H. Eberly, {\em Optics Letters} {\bf 36}, 4110 (2011).
\bibitem{q2}
X. F. Qian and J. H. Eberly, arXiv:1307.3772. 
\bibitem{aiello}
A. Aiello, F. T\"{o}ppel, C. Marquardt, E. Giacobino and G. Leuchs, arXiv: 1409.0213 [quant-ph] (2014).
\bibitem{pereira}
L. J. Pereira, A. Z. Khoury and K. Dechoum, arXiv: 1409.0889 [quant-ph] Sept (2014). 
\bibitem{K}
B. O. Koopman,  {\em Proc. Natl. Acad. Sci. U.S.A.} {\bf 17}, 315 (1931).
\bibitem{vN}
J. von Neumann,  {\em Ann. Math.} {\bf 33}, 587 (1932); {\em ibid.} {\bf 33}, 789-791 (1932).
\bibitem{kvn1}
E. Gozzi, {\em Phys. Lett. B} {\bf 201}, 525 (1988).
\bibitem{kvn2}
E. Gozzi, M. Reuter, and W. Thacker, {\em Phys. Rev. D} {\bf 40},
3363 (1989).
\bibitem{kvn3}
J. Wilkie and P. Brumer, {\em Phys. Rev. A} {\bf 55}, 27 (1997).
\bibitem{kvn4}
J. Wilkie and P. Brumer, {\em Phys. Rev. A} {\bf 55}, 43 (1997).
\bibitem{kvn5}
E. Gozzi and D. Mauro, {\em Ann. Phys.} {\bf 296}, 152 (2002).
\bibitem{mauro}
D. Mauro, Ph. D. thesis, arXiv:quant-ph/0301172.
\bibitem{kvn6}
E. Deotto, E. Gozzi, and D. Mauro, {\em J. Math. Phys.} {\bf 44},
5902 (2003).
\bibitem{kvn7}
E. Deotto, E. Gozzi, and D. Mauro, {\em J. Math. Phys.} {\bf 44},
5937 (2003).
\bibitem{kvn8}
A. A. Abrikosov, E. Gozzi, and D. Mauro, {\em Ann. Phys.}
{\bf 317}, 24 (2005).
\bibitem{kvn9}
M. Blasone, P. Jizba, and H. Kleinert, {\em Phys. Rev. A} {\bf 71},
052507 (2005).
\bibitem{kvn10}
P. Brumer and J. Gong, {\em Phys. Rev. A} {\bf 73}, 052109 (2006).
\bibitem{kvn11}
P. Carta, E. Gozzi, and D. Mauro, {\em Ann. Phys. (Leipzig)}
{\bf 15}, 177 (2006).
\bibitem{kvn12}
E. Gozzi and C. Pagani, {\em Phys. Rev. Lett.} {\bf 105}, 150604 (2010).
\bibitem{kvn13}
E. Gozzi and R. Penco, {\em Ann. Phys.} {\bf 326}, 876 (2011).
\bibitem{kvn14}
E. Cattaruzza, E. Gozzi, and A. F. Neto, {\em Ann. Phys.}
{\bf 326}, 2377 (2011).
\bibitem{sud}
E. C. G. Sudarshan, {\em Pramana} {\bf 6}, 117-126 (1976).
\bibitem{schmidt}
E. Schmidt, {\em Ann. Math.} {\bf 63}, 433-476 (1907).
\bibitem{bond}
D. I. Bondar, R. Cabrera, D. V. Zhdanov and H. A. Rabitz, {\em Phys. Rev.A} {\bf 88}, 052108 (2013); arXiv:quant.phys/1202.3628 (2012).
\bibitem{bond2}
D. I. Bondar, R. Cabrera, R. R. Lompay, M. Yu. Ivanov and H. A. Rabitz, {\em Phys. Rev. Lett.} {\bf 109},190403 (2012): arXiv:quant.phys/1105.4014 (2012).
\end{thebibliography}
\end{document}